\documentclass{ws-p9-75x6-50}
\begin{document}
\title{Axion in large extra dimension}
\author{Sanghyeon Chang, Shiro Tazawa and Masahiro Yamaguchi}
\address{Department of Physics, Tohoku University,
Sendai 980-8578, Japan}
\maketitle
\abstracts{
  We examine  the axion model  in the large extra dimensions scenario 
with TeV scale  gravity. 
To obtain an
  intermediate-scale decay constant of the axion, the axion is assumed
  to live in a sub-spacetime (brane) of the whole bulk. In this model
  there appear Kaluza-Klein modes of the axion which have stronger
  interaction than those of the graviton.  The axion brane plays
  a role of absorber of the graviton Kaluza-Klein modes. 
  We discuss various cosmological constraints as well as
  astrophysical ones and show that the model is viable for certain
  choices of the dimensionality of the axion brane. The structure of
  the model proposed here provides a viable realization of the fat
  brane idea to relax otherwise very severe cosmological constraints.
}

\section{Introduction}
It has been suggested that the fundamental scale of  nature can be
as low as TeV, whereas the largeness of the effective Planck scale or 
the weakness of the
gravity in a long distance can be explained by introducing large extra
dimensions\cite{ADD1,AADD,ADD2,ADM}.  When there exist $n$ of such extra 
dimensions, 
the relation between the gravitational scale
in $4$ dimensions and the fundamental scale 
in $(4+n)$ dimensions is given
\begin{equation}
M_{pl(4)}^2 \sim V_n M_{pl(4+n)}^{n+2} ,
\end{equation}
where $V_n$ is the volume of the extra dimensional space.
For $M_{pl(4+n)}\sim 1$ TeV, the size of the extra dimensions $r_n$ is 
computed as
\begin{equation}
r_n\sim V_n^{1/n}
\sim 10^{\frac{32}{n} -17} M_{\rm TeV}^{-\frac{2}{n} -1} {\rm cm} ,
\end{equation}
where $M_{\rm TeV} \equiv M_{pl(4+n)}/$TeV.
The case $n=1$ is excluded because the gravitational law would change at 
macroscopic level,  but the cases $n \geq 2$ are allowed by gravity 
experiments. 

Because there exist a tower of graviton Kaluza-Klein (KK) modes, 
various astrophysical and cosmological bounds will be applied\cite{ADD2,hall,photon}.
We may avoid some of these problems by assuming an extremely low reheating
temperature of the early universe. But the reheating temperature cannot
be smaller than $1$ MeV.

On the other hand, the simplest model of 
this type would not provide intermediate scales
which are necessary to explain phenomenological issues like, 
small neutrino masses
and the strong CP problem. It was pointed out\cite{ADD2} that
particles dwelling in the extra dimensions other than graviton can
also have similar effective interaction terms in four-dimensional
physics.  The interaction between bulk matter and normal
matters is suppressed as gravity.

In this paper we will propose an axion
model in TeV scale gravity for various numbers of the extra
dimensions, with special emphasis on cosmological constraints: the
addition of the axion in the large-extra-dimension model substantially
alters the cosmology.
 
We will use the convention
$M_{pl}\equiv M_{pl(4)}$ and $M_*\equiv M_{pl(4+n)}$.

\section{PQ scale in extra-dimension}

If an axion is a boundary field confined on
 4 dimensions, the Peccei-Quinn (PQ) scale $f_{PQ}$ is bounded by $M_* \sim
1$ TeV.  To obtain a higher PQ scale, the axion field has to be inevitably 
a bulk field. If it lives in the whole (4+n) dimensional bulk, the PQ
scale will be $f_{PQ} \sim M_{pl}$.

However, the damped coherent oscillation of the axion with 
$f_{PQ} \sim M_{pl}$ would overclose the universe. A conventional 
argument gives an upper bound of $F_{PQ}\sim 10^{12}$ GeV. 
Even when an entropy
production takes place after the QCD phase transition ({\it e.g.} the
reheating temperature is smaller than $\sim 1$ GeV), $f_{PQ}$ cannot be 
much larger than $10^{15}$ GeV\cite{kawa}.
Provided that the coherent
oscillation of the inflaton is followed by the reheating process, 
\begin{equation}
f_{PQ} < 10^{15} \mbox{GeV} \left(\frac{h}{0.7}\right)
\left(\frac{\pi/2}{\theta}\right)
\left(\frac{\mbox{MeV}}{T_R}\right)^{1/2},
\end{equation}
where $h$ is from Hubble constant in units of 100km\,sec$^{-1}$Mpc$^{-1}$, 
$\theta$ is the initial value of PQ vacuum angle and $T_R$ is the reheating
temperature after inflation.
To avoid over-closure problem, 
we propose  a natural way to realize an intermediate
scale axion, i.e. axion as as  $(4+m)$ dimensional sub-spacetime field ($(3+m)$
brane) ($m<n$). The idea of using
sub-spacetime to realize an intermediate scale already appeared in the context of neutrinos\cite{neutrino2}.

Let $\chi$ be a complex scalar field which contains PQ axion in 4+m
dimension; $\tilde{a}$.
If the axion field lives only on $4+m$ dimensional sub-spacetime 
where $m<n$ and  the volume of extra-dimension is $V_m$,
\begin{eqnarray}
{\cal L}_\chi =   \int dx^{4+m} \partial^M \chi^* \partial_M \chi
+ \int dx^4 \frac{\tilde{a}(x^A=0)}{\langle \chi\rangle} F\tilde{F}  
\end{eqnarray}
where $x^A$ means an extra-dimension coordinate. Assuming that the
vacuum expectation value of the $\chi$ field does not depend on the
extra-dimension coordinates, we obtain 
\begin{eqnarray}
f_{PQ}&\sim& \sqrt{V_m}\langle \chi \rangle \sim r_n^{m/2} M_*^{1+m/2}
\sim M_*\left(\frac{M_{pl}}{M_*}\right)^{m/n}
\nonumber \\
&\sim& 10^{3(1+5m/n)} M_{\rm TeV}^{1-m/n} \mbox{GeV} .
\end{eqnarray}
Here we have defined the 4-D axion field as $a= \sqrt{V_m}\tilde{a}(x^A=0)$ 
and assumed that the size $r_n$ is common for all extra dimensions.

A lower bound of $f_{PQ}$ comes from astrophysical bounds.
It is known that $f_{PQ}$ should be larger than $10^{9}$GeV.
In extra dimension physics, KK modes also contribute to supernova cooling
if their masses are smaller than the core temperature ($\sim 30$MeV) as
we will show shortly. 

\renewcommand{\arraystretch}{1.5}
\begin{table}
\caption{$f_{PQ}$ and lifetimes of axion and graviton KK modes
for $M_*=1$ TeV, where $M_{100}\equiv
\frac{m}{100\rm MeV}$}
\label{table1}
\footnotesize
\begin{center}
\begin{tabular}{|ccccc|}
\hline
$(m, n)$ &$r_n^{-1}[\mbox{MeV}]$& $f_{PQ}[\mbox{GeV}]$  
& $\tau_g[\mbox{sec}]$ & $\tau_a[\mbox{sec}]$   \\
\hline
$(1,2)$ &$4\times 10^{-10}$ &  $ 5\times 10^{10}$  & 
$2\times 10^{7}M_{100}^{-4}$ &
$6\times 10^{7}M_{100}^{-3}$\\
$(2,3)$ &$6\times 10^{-5}$ &  $2\times 10^{13}$       & 
$1\times 10^{6}M_{100}^{-5}$ &
$8\times 10^{12}M_{100}^{-3}$\\
$(2,4)$ &$2\times 10^{-2}$ &  $ 5\times 10^{10}$  & 
$2\times 10^{11}M_{100}^{-5}$ &
$6\times 10^{7}M_{100}^{-3}$\\
$(3,4)$ & &  $ 3\times 10^{14}$ & 
$3\times 10^{7}M_{100}^{-6}$ &
$3\times 10^{15}M_{100}^{-3}$\\
$(2,5)$ &$0.7$ &  $1\times 10^{9}$          & 
$2\times 10^{14}M_{100}^{-5}$ &
$5\times 10^{4}M_{100}^{-3}$\\
$(3,5)$ & &  $2\times 10^{12}$       & 
$1\times 10^{12}M_{100}^{-6}$ &
$7\times 10^{10}M_{100}^{-3}$\\
$(4,5)$ & &  $2\times 10^{15}$       & 
$9\times 10^{9}M_{100}^{-7}$ &
$1\times 10^{17}M_{100}^{-3}$\\
$(3,6)$ &$7$ &  $ 5\times 10^{10}$  & 
$2\times 10^{15}M_{100}^{-6}$ &
$6\times 10^{7}M_{100}^{-3}$\\
$(4,6)$ & &  $2\times 10^{13}$       & 
$1\times 10^{14}M_{100}^{-7}$ &
$8\times 10^{12}M_{100}^{-3}$\\
\hline
\end{tabular}
\end{center}
\end{table}

\begin{table}
\caption{$f_{PQ}$ and lifetimes of axion and graviton KK modes
for $M_*=10$ TeV.}
\label{table2}
\begin{center}
\footnotesize
\begin{tabular}{|ccccc|}
\hline
$(m, n)$ &$r_n^{-1}[\mbox{MeV}]$ & $f_{PQ}[\mbox{GeV}]$  & $\tau_g[\mbox{sec}]$ & $\tau_a[\mbox{sec}]$   \\
\hline
$(1,2)$ &$4\times 10^{-8}$ &  $2\times 10^{11}$  & 
$2\times 10^{9}M_{100}^{-4}$ &
$6\times 10^{8}M_{100}^{-3}$\\
$(2,3)$ &$3\times 10^{-3}$ &  $4\times 10^{13}$       & 
$3\times 10^{9}M_{100}^{-5}$ &
$4\times 10^{13}M_{100}^{-3}$\\
$(2,4)$ &$0.6$ &  $2\times 10^{11}$  & 
$2\times 10^{14}M_{100}^{-5}$ &
$6\times 10^{8}M_{100}^{-3}$\\
$(3,4)$ & &  $6\times 10^{14}$ & 
$1\times 10^{12}M_{100}^{-6}$ &
$9\times 10^{15}M_{100}^{-3}$\\
$(2,5)$ &$20$ &  $6\times 10^9$          & 
$1\times 10^{17}M_{100}^{-5}$ &
$8\times 10^{5}M_{100}^{-3}$\\
$(3,5)$ & &  $ 4\times 10^{12}$       & 
$2\times 10^{16}M_{100}^{-6}$ &
$5\times 10^{11}M_{100}^{-3}$\\
$(4,5)$ & &  $3\times 10^{15}$       & 
$4\times 10^{15}M_{100}^{-7}$ &
$3\times 10^{17}M_{100}^{-3}$\\
$(3,6)$ &$160$ &  $2\times 10^{11}$  & 
$2\times 10^{19}M_{100}^{-6}$ &
$6\times 10^{8}M_{100}^{-3}$\\
$(4,6)$ & &  $4\times 10^{13}$       & 
$3\times 10^{19}M_{100}^{-7}$ &
$4\times 10^{13}M_{100}^{-3}$\\
\hline
\end{tabular}
\end{center}
\end{table}
To have $10^{9}$ GeV $<f_{PQ} \leq 10^{15}$ GeV, we need $2/5<m/n \leq
4/5$.  Possible sets of ($m,n$) with $f_{PQ}$ where $M_*= 1$ TeV and
$10$ TeV can be found in tables.

\section{Laboratory and Astrophysical constraints}
The strongest bound from astrophysics is the bound from the Supernova cooling.
In SN1987A observation, it was calculated that 
\begin{equation}
\frac{1}{f_{PQ}^2} < 10^{-18} \mbox{ GeV}^{-2},
\end{equation}
Since the axion KK modes interact exactly the same way as the conventional 
axion, 
the effective interaction of the KK modes at the
core temperature $T\simeq 30$ MeV is
\begin{equation}
\frac{1}{f_{PQ}^2}\times (T r_n)^m \sim \frac{T^m}{M_*^{m+2}} <10^{-18}
\mbox{ GeV}^{-2}.
\end{equation}
This gives a bound on the fundamental scale
\begin{equation}
M_* > (10^{18} \times 0.03^m)^\frac{1}{m+2} \mbox{ GeV}.
\end{equation}
For $m=1$, $M_*>300$ TeV and $m=2$, $M_* > 5$ TeV. $M_* \sim 1$ TeV is allowed
only if $m>2$.

In near future, high energy accelerator experiments may probe 
the axion KK mode emission.   The graviton KK mode signals in collider
were discussed in detail recently\cite{han2}.
The scattering cross section of the graviton KK mode emission from high energy 
scattering with center of mass frame energy $\sqrt{s}$ is 
\begin{equation}
\sigma \propto \frac{1}{M_{pl}^2} (\sqrt{s}r_n)^n
\sim \left(\frac{\sqrt{s}}{M_*}\right)^{n+2}  \frac{1}{s},
\end{equation}
while the KK axion production is
\begin{equation}
\sigma \propto \frac{1}{f_{PQ}^2} (\sqrt{s}r_n)^m
\sim \left(\frac{\sqrt{s}}{M_*}\right)^{m+2} \frac{1}{s}.
\end{equation}
Since the energy dependence of the axion KK mode
cross-section is different from the
graviton KK mode cross-section,
it might be possible to detect 
this difference at TeV scale collider experiments. 

\section{Thermal production of Axion KK mode}
The axion KK mode masses are proportional to $r_n^{-1}$ and
they have stronger couplings than the graviton KK modes to the matters in our
universe in general cases.
If there is no hidden particle which couples to the axion, 
the main decay channel of rather light  KK axion is to two photons.
\begin{equation}
 \Gamma_{a_{KK}\rightarrow 2\gamma} \simeq \frac{C_{a\gamma}^2}{64\pi} 
\left(\frac{\alpha}{\pi}\right)^2 \frac{m_{A}^3}
{f_{PQ}^2} \sim 3\cdot10^{-8} C_{a\gamma}^2  \frac{m_{A}^3}{f_{PQ}^2},
\end{equation}
where $m_A$ is the mass of the axion KK mode and
$C_{a\gamma}$ is the axion-photon coupling
which is usually within $0.1$ to $1$.
This decay can be  cosmologically dangerous.
For instance, for $f_{PQ}= 10^{12}$ GeV and $m_{A}=1$ MeV,
 life time of KK mode is $\tau_{A} \sim 10^{17}$ sec, which is about the age of the universe.

The graviton KK modes have similar cosmological problems
because they can overclose our universe or decay into photons at a
late stage of cosmological evolution.  It was suggested
that a ``fat brane"\cite{ADD2} can solve the cosmological problems by
absorbing most of the decay products of the KK modes.  However
massless particles in the higher dimensional brane {\it are not}
massless in our four-dimensional universe, 
thus it cannot solve over-closure problem\cite{hall}.

If we add an axion as a ``brane particle", the graviton KK mode
will decay to the ``brane" axion more efficiently, since its decay
width is enhanced by factor $(m r_n)^m$.
Because the massive axion KK mode can decay into the photon pairs stronger
than graviton KK mode, 
the primordial graviton KK mode
will not over-close the universe. Instead, it will contribute 
to the cosmological background radiation. 
Axions can be produced thermally during the reheating process, which 
is also much stronger process than graviton.
This can be a severe constraint to the axion model. 

In our longer paper\cite{CTY} the yields $Y$ 
from four different sources of the axion KK modes were calculated, but here
we will present only two relevant sources:\\ 
{\bf I.} the axion KK mode from the pion scattering ($\pi\pi \rightarrow \pi
a_{KK}$); this process dominates if $T_R > 10$ MeV \\
{\bf II.} the axion KK mode from the two photon inverse decay  ($2\gamma \rightarrow
a_{KK}$); this gives significant contribution when $m_{A}\sim T_R$.
 
\begin{eqnarray}
Y_I &\simeq& 6\cdot 10^{-10} \left(\frac{10^{12}\rm GeV}{f_{PQ}}\right)^2
\left(\frac{T_R}{100\mbox{MeV}}\right)^3 A_I,
\end{eqnarray}
\begin{eqnarray}
A_I&=& C_{a\pi}^2\left(\frac{10}{g_*(T_R)} \right)^{3/2} 
\left(\frac{I(T_R)}{1000}\right)
\end{eqnarray}
\begin{eqnarray}
Y_{II} &\simeq& 2\cdot 10^{-16} \left(\frac{10^{12}\rm GeV}{f_{PQ}}\right)^2 
\left(\frac{ T_R}{100\rm MeV}\right) 
C_{a\gamma}^2\left(\frac{10}{g_*(T_R)} \right)^{3/2}
\left(\frac{m_{A}}{T_R} \right)^3.
\end{eqnarray}
For the details of these calculations and the definition of
function $I(T)$, see ref.\cite{CTY}.
The lifetime of each KK mode for given $(n,m)$ can be found in tables.
\begin{figure}
\vspace{-5mm}
\begin{center}
\epsfxsize 10cm
\epsfbox{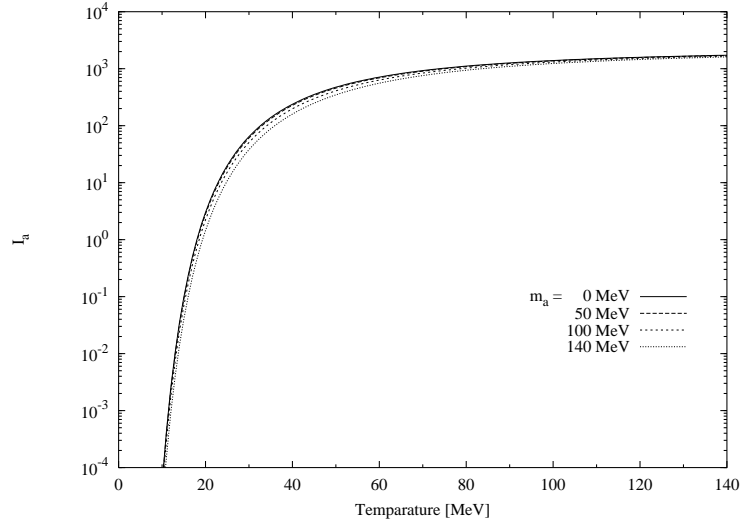}
\end{center}
\caption{The function $I(T)$ for various $m_{A}$ in the KK mode yield $Y_I$.}
\end{figure}

\section{Cosmological constraints}
In table 1 and 2, $n\geq 5$ in both $M_*=1 $ and
$10$ TeV cases and $n=4$ in $1$ TeV have
  minimal KK mode mass
greater than $1$ MeV. If one takes the minimal reheating temperature
of 1 MeV, these cases are cosmologically safe.
On the other hand $m \leq 2$ in $M_*=1 $ TeV and $m=1$ in $M_*=10$ TeV
is forbidden by the astrophysical bound.
The cases $n=4, m=3$ at $M_*=1$ TeV and  $n=3, m=2$ at $M_*=10$ TeV are
not trivially safe or ruled out by the cosmological constraints.
\subsection{Big bang nucleosynthesis}
At the temperature of the universe around 1 MeV, it is required that
there should not be additional particles 
 which contribute to
the energy density significantly. Otherwise $^4$He would be produced
more than what is observed.  

The yield of the axion KK mode at BBN for $T_R>10$ MeV is
\begin{eqnarray}
\left.\frac{\rho_A}{s}\right|_{BBN} &\simeq &
 m_1 Y_I (m_1 r_n)^m
 < 0.1 \mbox{ MeV} ,
\end{eqnarray}
KK mode may practically have maximal mass 
$m_1\equiv \max\{m_\pi, T_R\} $
which is produced in the thermal bath.

If $T_R\sim m_\pi$, approximately 
\begin{equation}
M_*>10^{\frac{20}{m+2}} \times m_\pi.
\end{equation}
For  $m=1$, this reads $M_* > 600$ TeV, and for $m=2$, $M_* >15 $ TeV.
But if the reheating temperature is as low as $10$ MeV, this bound 
is not important since $A_I(T_R)$ is suppressed exponentially.

\subsection{Over-closure of Universe}
The total energy of the axion KK modes at present must not exceed the
critical density:
\begin{equation}
\rho_A <\rho_c = 3\cdot10^{-6} s_0 h^2 \mbox{ MeV} ,
\end{equation}
where $s_0\simeq 3000$cm$^{-3}$ is the entropy of the present universe.

For the case that the KK modes decay into some relativistic particles,
we can divide the bound in two parts; decay before the present
time and do not decay till now:
\begin{eqnarray}
\frac{\rho_A}{s_0}&\sim& \int^{m_2}_{m_0} dm_A (m_A r_n)^m Y_A
+\int^{m_1}_{m_2} dm_A (m_A r_n)^m \frac{Y_A T_0}{T(m_A)}
\end{eqnarray}
where the axion KK mode with mass $m_2$ decays at the present time.

\subsection{Cosmological Microwave Background Radiation}
If the massive KK modes decay after 
$10^6$ sec but before the recombination era, the produced
photons may give a distortion of the cosmological microwave background
radiation. The COBE observation gives a bound\cite{CMBR}
\begin{equation}
\frac{\Delta \rho_\gamma}{s} \leq 2.5 \times 10^{-5} T_D
\end{equation}
where $T_D$ is temperature at KK mode decay.

\subsection{Diffuse photon background}
Observations of diffuse photon backgrounds at the present universe give
upper bounds on additional contributions to photon spectrum. For example,
for the energy range $800$ keV $<E <30$ MeV\cite{diffuse-photon}
\begin{equation}
\frac{d{\cal F}}{d\Omega} 
< 78 \left(\frac{E}{ 1 {\rm  keV}}\right)^{-1.4}.
\end{equation}
Constraints on other ranges of the photon energy can be found, e.g. in 
Ref.\cite{KY}

Theoretical prediction is
\begin{equation}
\frac{d{\cal F}}{d\Omega} = \frac {n_{A} c}{4\pi} \times Br
\end{equation}
for the life-time of KK mode shorter than the age of the universe, and
\begin{equation}
\frac{d{\cal F}}{d\Omega} \sim Br \times\frac {n_{A} c}{4\pi}
\frac{\Gamma_{a_{KK}\rightarrow
2\gamma} }{Br H_0} \left( \frac{2E}{m_Ac^2}\right)^{3/2} (m_Ar_n)^m
\end{equation}
for its life time  longer than the age of the universe\cite{Kolb-Turner}.
$Br$ is a branching ratio of axion decay into two photons.

Since the axion decay width is highly suppressed
by $\left(\frac{\alpha}{\pi}\right)^2$ with $\alpha$ being the fine structure
constant
it is easy to get a low branching ratio to decay into the photon. 
We can assume that there exist another 4D  wall in the brane. Or
one can just imagine that there are some unknown particles on our wall.
Most of the axions decay into the other wall if the
coupling constant, the color factor of the other gauge interaction,
and/or the number of the fermions with PQ charges in the decay loop
diagram in the other wall (or the other particles) are large enough.

\subsection{Results}

{\bf I.  $n=4,\ m=3$ and $M_*=1$ TeV case:} \\
1. BBN bound 
\begin{equation}
T_R < 80  \mbox{ MeV}
\end{equation}
2. Over-closure bound,
\begin{equation}
T_R <  12\, 
\left(\frac{C_{a\gamma}^2}{A_I^2 Br}\right)^{\frac{1}{6}} 
\mbox{ MeV}
\simeq 30 \mbox{ MeV  (for } Br=1) .
\end{equation}
3. CMBR bound, (for $m_A\geq 100$ MeV and $T_R>10$ MeV)
\begin{equation}
T_R < 2 \cdot 10^{-2} 
\left(\frac{C_{a\gamma}^2}{A_I^2 Br^3}\right)^{\frac{1}{6}} 
\mbox{ MeV}
\end{equation}
4. Diffused photon bound, \\
for $T_R> 10$ MeV, 
\begin{equation}
T_R< 2\cdot 10^{-2} \left(\frac{Br}{C_{a\gamma}}\right)^{-0.63} \left(
\frac{m_A}{10\rm MeV}\right)^{-0.53}A_I^{-1/3} \mbox{ MeV},
\end{equation}
for $T_R< 10$ MeV, 
\begin{equation}
T_R< 0.3\, Br^{-0.73} \mbox{ MeV}.
\end{equation}
for $Br<\Gamma_{a\rightarrow 2\gamma}/H_0$, and 
\begin{equation}
T_R < 5\, C_{a\gamma}^{-0.48} \mbox{ MeV}
\end{equation}
where $Br > \Gamma_{a\rightarrow 2\gamma}/H_0$.

\noindent{\bf II. $n=3,\ m=2$ and $M_*=10$ TeV case:} \\
1. BBN bound 
\begin{equation}
T_R <  90 \mbox{ MeV}.
\end{equation}
2. Over-closure bound,
\begin{equation}
T_R <  28\, 
\left(\frac{C_{a\gamma}^2}{A_I^2 Br}\right)^{\frac{1}{6}} \mbox{ MeV} \simeq
40 \mbox{ MeV (for } Br=1).
\end{equation}
3. CMBR bound,
\begin{equation}
T_R < 4 \cdot 10^{-2} 
\left(\frac{C_{a\gamma}^2}{A_I^2 Br^3}\right)^{\frac{1}{6}} 
\mbox{ MeV}.
\end{equation}
4. Diffused photon bound,\\
for $T_R> 10$ MeV, 
\begin{equation}
T_R< 3\cdot 10^{-2} \left(\frac{Br}{C_{a\gamma}}\right)^{-0.63} \left(
\frac{m_A}{10\rm MeV}\right)^{-0.2}A_I^{-1/3} \mbox{ MeV},
\end{equation}
for $T_R< 10$ MeV, 
\begin{equation}
T_R< 0.1\, Br^{-1.2} \mbox{ MeV},
\end{equation}
for $Br<\Gamma_{a\rightarrow 2\gamma}/H_0$, and 
\begin{equation}
T_R < 3 \mbox{ MeV}
\end{equation}
where $Br > \Gamma_{a\rightarrow 2\gamma}/H_0$.

\begin{figure}
\vspace{-5mm}
\begin{center}
\epsfxsize 9cm
\epsfbox{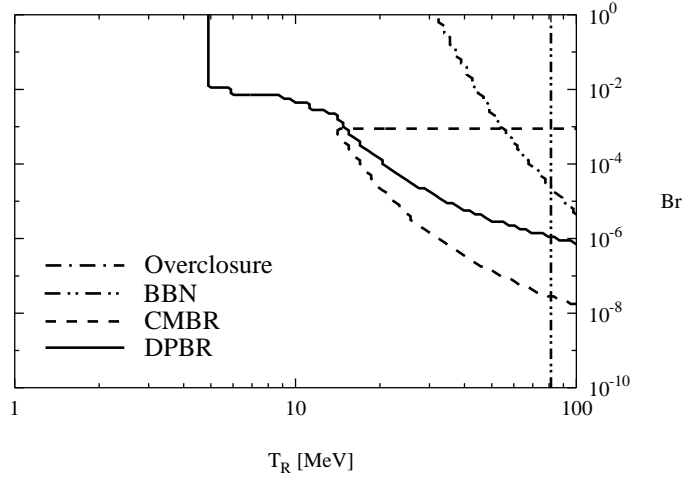}
\end{center}
\caption{Bound for $M_*=1$ TeV and $n=4, m=3$,
Upper and right side of each line is excluded region.}
\end{figure}

\begin{figure}
\vspace{-5mm}
\begin{center}
\epsfxsize 9cm
\epsfbox{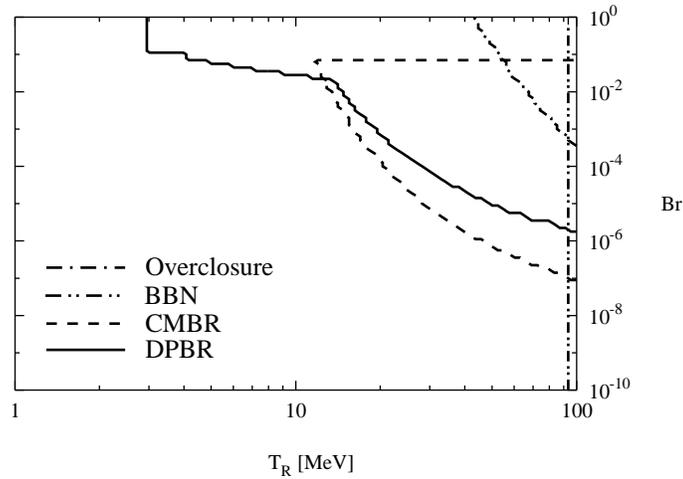}
\end{center}
\caption{Bound for $M_*=10$ TeV and $n=3, m=2$. The
excluded region is same as Fig.~2.}
\end{figure}

We also performed computer calculations on both $n=4,\ m=3$ and
$M_*=1$ TeV
and $n=3,\ m=2$ and $M_*=10$ TeV cases. Fig. 2 and 3 show that
the DPBR is the stringent bound if the branching ratio is small. But in an
extremely small
branching ratio case, CMBR is dominant. This is because the life time of the
axion KK modes becomes shorter than $10^{13}$ sec, and so the produced photons
will disturb the CMBR spectrum.  

We also performed full computer calculations for all possible $(m,n)$ sets
in tables in our longer paper\cite{CTY}.

\section{conclusions}

In this paper, we discussed the axion model in the  extra dimensions whose 
PQ scale lies in an
intermediate scale $f_{PQ} \leq 10^{15}$ GeV. This intermediate scale can
be obtained by introducing a $3+m$ dimensional brane in the 
$4+n$ dimension bulk.

If we include the axion as a brane particle, it will change 
the extra dimension physics, especially in the cosmological aspect.
Since the graviton KK mode will decay into the axion KK mode, the
over-closure problem is not as serious as the original model of Arkani-Hamed
et.al. 
On the other hand, the argument from stars and supernova cooling will
give a more strict bound on the axion production. 
Among other things, the most sever cosmological bound comes from photon 
emission through the decays of the KK modes of the axion. 
We found that the astrophysical argument restrict the number of the dimensionality of the sub-spacetime where the axion lives: $m>2$ for $M_*=1$ TeV and 
$m>1$ for $M_*=10$ TeV.
The latter case requires quite a low reheating temperature
after the inflation.

To lift this bound, we can introduce the hidden matter/gauge fields to another four-dimensional wall (or to our wall itself) which has much stronger coupling to
axion and/or much more generations of particles, (or maybe much lower
QCD phase transition scale, etc). This can significantly lower the
branch ratio of the axion KK mode decay into photons.

This idea can be used to improve the original {\em fat brane} model. 
The intermediate scale brane can absorb most of graviton KK modes, and
the particle in this brane can decay
into relativistic particles in the four dimensional wall(s). 
This mechanism can solve the problem of 
the overclosure of the universe by the KK modes. If  
we introduce, in this brane,  a massless particle (which is not an axion) which
does not decay into photons or any dangerous particles in 4-D, 
we can avoid other cosmological problems such as the 
ones related to the cosmic photon backgrounds.

\end{document}